\documentclass[aps,prd,superscriptaddress,showpacs,nofootinbib,notitlepage,twocolumn]{revtex4-1}
\usepackage{graphicx,subfigure,amsmath,amsfonts,amssymb,multirow}
\usepackage[colorlinks,linkcolor=blue,citecolor=blue,urlcolor=blue]{hyperref}

\def\nat{Nature\ }
\def\aap{Astron.\ Astrophys.\ }
\def\apj{Astrophys.\ J.\ }
\def\apjl{Astrophys.\ J.\ Lett.\ }
\def\apjs{Astrophys.\ J.\ Supp.\ }

\def\mnras{Mon.\ Not.\ Roy.\ Astron.\ Soc.\ }
\def\physrep{Phys.\ Rept.\ }
\def\prd{Phys.\ Rev.\ D\ }
\def\prl{Phys.\ Rev.\ Lett.\ }

\def\araa{Annu.\ Rev.\ Astron.\ Astrophys.\ }
\def\nphysa{Nucl.\ Phys.\ A\ }

\begin{document}
\newcommand{\PMO}{Key Laboratory of Dark Matter and Space Astronomy, Purple Mountain Observatory, Chinese Academy of Sciences, Nanjing, 210033, People's Republic of China.}
\newcommand{\USTC}{School of Astronomy and Space Science, University of Science and Technology of China, Hefei, Anhui 230026, People's Republic of China.}
\newcommand{\NJNU}{Department of Physics and Institute of Theoretical Physics, Nanjing Normal University, Nanjing 210046, People's Republic of China.}

\title{Constraint on phase transition with the multimessenger data of neutron stars}
\author{Shao-Peng Tang}
\author{Jin-Liang Jiang}
\affiliation{\PMO}
\affiliation{\USTC}
\author{Wei-Hong Gao}
\affiliation{\NJNU}
\author{Yi-Zhong Fan}
\email[Corresponding author.~]{yzfan@pmo.ac.cn}
\author{Da-Ming Wei}
\affiliation{\PMO}
\affiliation{\USTC}
\date{\today} 

\begin{abstract}
The equation of state (EoS) of the neutron star (NS) matter remains an enigma. In this work we perform the Bayesian parameter inference with the gravitational wave data (GW170817) and mass-radius observations of some NSs ({PSR J0030+0451}, {PSR J0437-4715}, and {4U 1702-429}) using the phenomenologically constructed EoS models to search for a potential first-order phase transition. Our phenomenological EoS models take the advantages of current widely used parametrizing methods, which are flexible enough to resemble various theoretical EoS models. We find that the current observation data are still not informative enough to support/rule out phase transition, due to the comparable evidences for models with and without phase transition. However, the bulk properties of the canonical $1.4\,M_\odot$ NS and the pressure at around $2\rho_{\rm sat}$ are well constrained by the data, where $\rho_{\rm sat}$ is the nuclear saturation density. Moreover, strong phase transition at low densities is disfavored, and the $1\sigma$ lower bound of transition density is constrained to $1.84\rho_{\rm sat}$.
\end{abstract}
\maketitle

\section{Introduction}

Neutron stars (NSs) are natural laboratories used to examine the unknown equation of state (EoS) of dense matter with the highest density in the Universe, which is mainly composed of hadrons or deconfined quark matter \citep{2007PrPNP..59...94W}. The EoS of the NS matter has been widely studied by theoretical calculations \citep{2009RvMP...81.1773E}, phenomenological parametrizations \citep{2009PhRvD..79l4032R, 2014PhRvD..89f4003L}, and nonparametric methods \citep{2020PhRvD.101f3007E}. The joint analyses of the $M-R$ measurements, gravitational wave (GW) data, and the observed maximum mass of NSs have set stringent constraints on the EoS \citep{2019ApJ...885...39J, 2020ApJ...892...55J, 2020PhRvD.101l3007L, 2020ApJ...888...12M, 2020ApJ...893L..21R, 2020arXiv200801582B, 2020PhRvC.102e5801K}. 

The nature of matter in the core of NSs remains to be better understood. Depending on the possible compositions of the unknown matter, the compact stars can be either normal NSs \citep{1939PhRv...55..374O, 1939PhRv...55..726Z}, hybrid stars \citep{2005ApJ...629..969A, 2013PhRvD..88h3013A}, or strange quark stars \citep{1986ApJ...310..261A, 1986A&A...160..121H}. At sufficiently high energy density, fundamental theories like quantum chromodynamics predict a deconfinement transition of hadronic nuclear matter into a new phase of quarks and gluons \citep{1980PhR....61...71S}. However, it is still not clear whether quark matter exists in NSs (or their merger remnants), or where/when the potential hadron-quark phase transition happens. By evaluating the sound velocity in strongly interacting matter, \citet{2020NatPh..16..907A} found that a sizable quark core in the massive NSs ($>2\,M_\odot$) should be present unless the conformal bound has been seriously violated. Many works have analyzed the feasibility of observing the presence of this exotic core (e.g., Refs.~\citep{2019AIPC.2127b0006H, 2020ApJ...893L...4C, 2020PhRvD.102f3003D}). The merger remnant of a binary neutron star, which is expected to have extremely high density after the collision, is especially promising for exhibiting a strong phase transition \citep{2019PhRvL.122f1101M, 2020PhRvD.102l3023B} that may leave imprints on the post-merger GW signals \citep{2018PhRvL.120z1103M, 2020PhRvL.124q1103W, 2020PhRvD.101j3006E}. While in the scenario that phase transition occurs at relatively low density, a particular family of NSs named ``twin stars," whose $M-R$ curve contains two or more branches, can be present when the transition is sufficiently strong \citep{1998PhRvD..58b4008L, 2000A&A...353L...9G, 2015A&A...577A..40B}. Therefore the $M-R$ characteristics can be adopted to constrain the hadron-quark phase transition \citep{2019arXiv190602522B, 2020ApJ...894L...8C}. Besides, the observed/inferred maximum mass of NSs (e.g., Refs.~\citep{2020NatAs...4...72C, 2018MNRAS.478.1377A}) may provide another ingredient to the research \citep{2007Natur.445E...7A, 2013A&A...553A..22C, 2014PhRvC..89a5806O, 2021ApJ...908..122G}. Benefiting from terrestrial nuclear experiments, $M-R$ measurements, and GW observations, numerous efforts (e.g., Refs.~\citep{2018RPPh...81e6902B, 2019PhRvD..99j3009M, 2019PhRvD..99b3009C, 2019PhRvD..99h3014H, 2019JPhG...46g3002O, 2020PhRvD.101d4019C, 2020ApJ...899..164H, 2020PhRvC.101b5807S}) have been made in the high-energy nuclear physics and astrophysics communities.

Previous studies about the nature of this transition were carried out based on various state-of-the-art theories for both the hadronic and quark phases. However, most of them used the traditional forward-modeling approach rather than Bayesian analysis. Usually it is difficult to quantify the uncertainties of model parameters and the evidence for presenting phase transition involved in forward-modeling methods. Therefore, to overcome these drawbacks, we analyze the observation data within the Bayesian statistical framework. Besides, phenomenologically parametrized EoS models can cover wider parameter space, and are more generically able to be incorporated in the Bayesian inference. Different phenomenological models, such as the piecewise polytrope \citep{2009PhRvD..79l4032R, 2009PhRvD..80j3003O}, the spectral representation \citep{2014PhRvD..89f4003L, 2018PhRvD..97l3019L}, and the constant-speed-of-sound (CSS) parametrization \citep{2013PhRvD..88h3013A}, have their own advantages on tackling the EoS with phase transition. However, it is rather challenging to mimic well all kinds of EoS models in the whole density range with a specific parametrization alone. In different density ranges, different parametrization approaches may be needed to better resemble the EoS model. For example, below the nuclear saturation density the EoS can be reasonably described by the piecewise polytrope. Between the nuclear saturation density and the phase transition onset density, spectral representation is more capable of capturing the features of theoretical EoSs. At even higher densities, the CSS may be a good approach. In this work, we aim to construct a generic phenomenological parametrization model that is flexible to resemble various theoretical EoS models and simultaneously fit the GW data (GW170817) and $M-R$ measurements of three NSs with the Bayesian inference method. Our main results show that current observation data are still not informative enough to support/rule out phase transition, but strong phase transition at low densities is disfavored and the bulk properties of canonical NSs are well constrained.

The elements for Bayesian inference, i.e., the parametrized EoS models, priors, and observation data, are respectively described in Secs.~\ref{sec:model}-\ref{sec:data}. We will introduce the Bayesian inference method for a joint analysis of GW data and the $M-R$ measurements in Sec.~\ref{sec:bayesian}, and present our main results in Sec.~\ref{sec:results}. Finally, Sec.~\ref{sec:summary} is our summary and discussion. Throughout this work, the uncertainties are for a 90\% confidence level unless specifically noticed.

\section{Parametrizing EoS}\label{sec:model}

We use a combination of a piecewise polytrope \citep{2009PhRvD..79l4032R, 2009PhRvD..80j3003O}, causal spectral representation \citep{2018PhRvD..97l3019L}, and constant-speed-of-sound parametrization \citep{2013PhRvD..88h3013A} to describe the EoS with a potential first-order transition between the hadronic and the quark phases. The adiabatic indices used to construct the EoS are expressed as
\begin{equation}\label{eq:Gamma}
    \Gamma(e,p,h) = \begin{cases}
                        \log{\frac{p_1}{p_0}}/\log{\frac{\rho_1}{\rho_0}} & \quad \rho_0<\rho \leq \rho_1, \\
                        \frac{1}{1+\Upsilon(h,v_k)}\frac{e+p}{p} & \quad \rho_1<\rho \leq \rho_2, \\
                        \Gamma_{\rm m} & \quad \rho_2<\rho \leq \rho_2\!+\!\Delta \rho, \\
                        c_{\rm q}^2\frac{e+p}{p} & \quad \rho>\rho_2\!+\!\Delta \rho,
                    \end{cases}
\end{equation}
where $e$, $p$, and $h$ denote, respectively, the internal energy density (including the rest mass contribution), the total pressure, and the pseudo enthalpy defined by $h(p)=\int_0^p\mathop{}\!\mathrm{d}p^\prime/[e(p^\prime)+p^\prime]$. Depending on the rest-mass density $\rho=(e+p)/\exp{\!(h)}$, the EoS can be divided into five segments. The first very low-density range of our model is interpolated by EoS SLy \citep{2001A&A...380..151D}, which determines the pressure $p_0$ at a density of $\rho_0=\rho_{\rm sat}/3$. In the second segment, we fix $\rho_1$ to $\rho_{\rm sat}$ and allow $p_1$ to vary in the range of $[3.12, 4.70]\times10^{33}\,{\rm dyn\,cm^{-2}}$ because the pressure at $\rho_{\rm sat}$ is well constrained by the nuclear theories/experiments \citep{2013ApJ...771...51L, 2017ApJ...848..105T}. Between the saturation density $\rho_{\rm sat}$ and the dividing density $\rho_2$, we adopt causal spectral representation \citep{2018PhRvD..97l3019L} to construct the EoSs. This method has additional freedom to more closely mimic some widely used EoS models. Especially when the potential complex behavior of hyperons dominates the density regime of $\rho_1<\rho \leq \rho_2$, using a constant adiabatic index to describe the EoS may be oversimplified.\footnote{It is expected that hyperons appear at around $2\rho_{\rm sat}$ and remain present in the dense matter until they are suppressed by quark deconfinement. For this reason, if the threshold density for the phase transition is above this value, the adiabatic index or speed of sound is expected to not have a constant slope.} The expression of $\Upsilon(h,v_k)$ is
\begin{equation}\label{eq:Upsilon}
    \Upsilon(h,v_k)=\exp \left\{ \sum_{k=0}^2 v_k \left[ \log{\left( \frac{h}{h_{\rm ref}}\right)} \right]^k \right\},
\end{equation}
where $v_k$ are the expansion coefficients and $h_{\rm ref}$ is the pseudo enthalpy at a density of $\rho_1$. While in the high-density region, the EoS is phenomenologically parametrized by the adiabatic index $\Gamma_{\rm m}$, the dividing density $\rho_2$, and the density jump $\Delta \rho$. Since the mixed phase is quite uncertain \citep{2018PhRvC..97d5802A}, we limit $\Gamma_{\rm m}$ in the range of $[0.01, 1.4]$ for the model of phase transition (PT) taking place in the nonrotating NS and the range of $[1.4, 10]$ otherwise [i.e., there is no phase transition (NPT)]. Thus in the PT model, $\rho_2$ represents the phase transition onset density. Above the density $\rho_2\!+\!\Delta \rho$, we use the CSS parametrization to describe the quark phase with the squared speed of sound parameter $c_{\rm q}^2$. We implement our model with the enthalpy-based formulae of \citet{2014PhRvD..89f4003L} to solve the Tolman-Oppenhimer-Volkoff and Regge-Wheeler equations (see also Refs.~\citep{2019PhRvD..99h3014H, 2020ApJ...888...45T}). Given the central conditions, like the energy density in the core, each possible EoS can uniquely determine the global structures of nonrotating NSs. Thus we can map the mass or central enthalpy to other bulk properties, e.g., radius $R$ and tidal deformability $\Lambda$.

\section{Constraints and Priors}\label{sec:prior}

\begin{figure*}
    \centering
    \includegraphics[width=0.98\columnwidth]{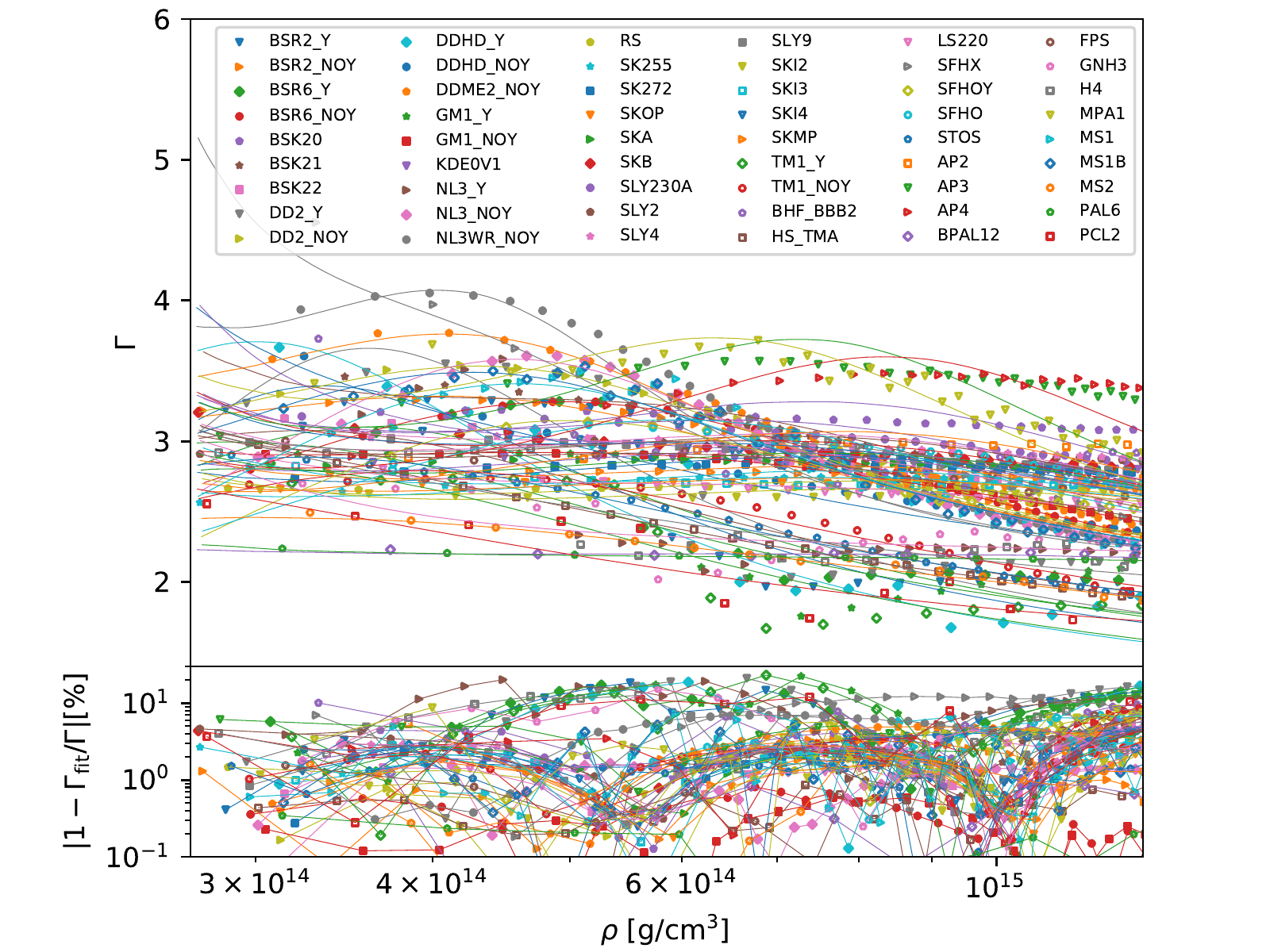}
    \includegraphics[width=0.98\columnwidth]{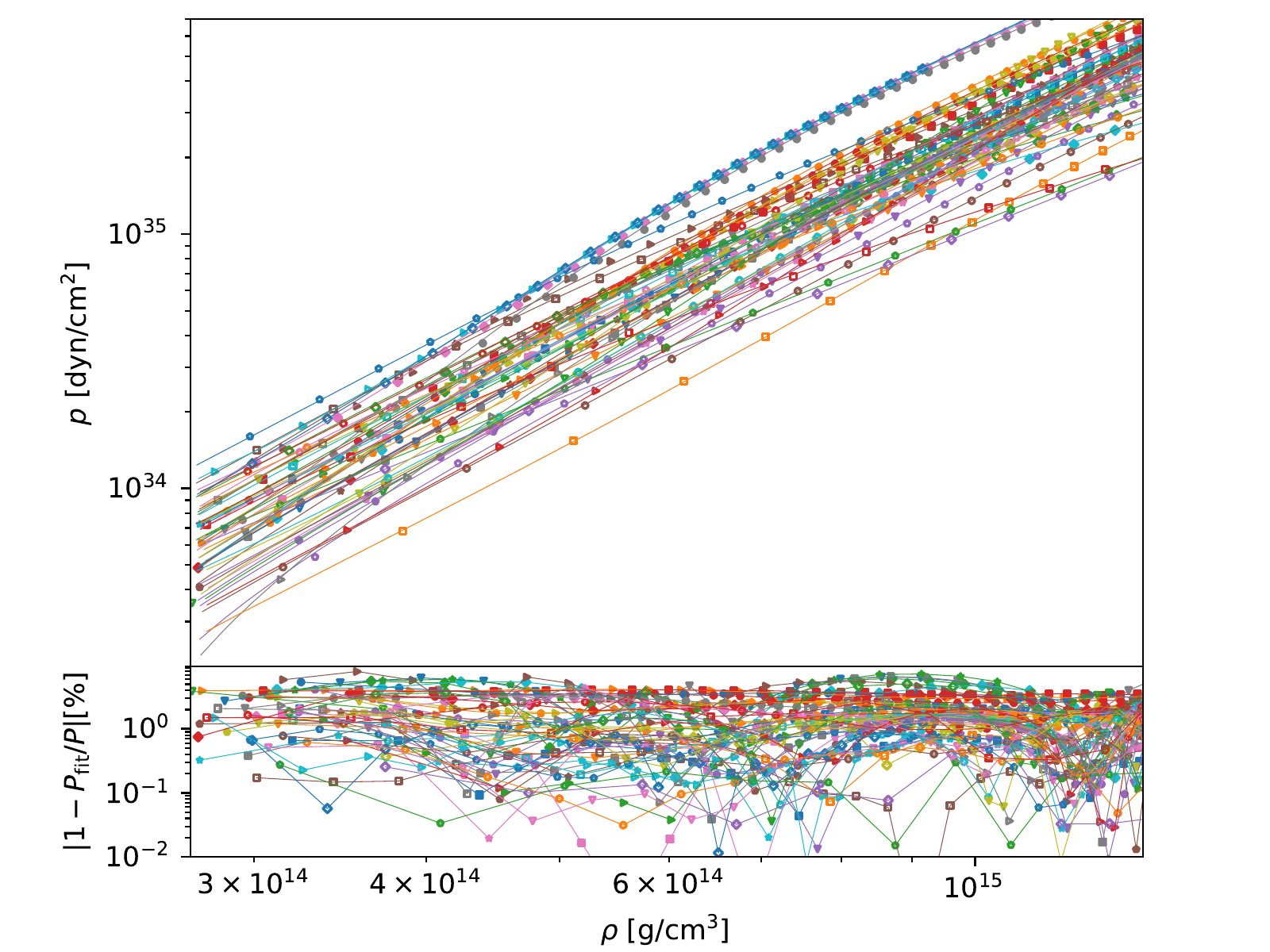}
    \caption{Fitting results of the rest-mass density versus the adiabatic index (left panel) and the rest-mass density versus the pressure (right panel) relations of some widely-studied/adopted theoretical EoSs using the causal spectral representation. In the top panels, the scatter points with different markers represent the data given by the EoS tables, and the solid lines are the corresponding best fit result for each EoS. The relative fitting errors are shown in the bottom panels.}
    \label{fig:fiteos}
    \hfill
\end{figure*}

\begin{figure*}
    \centering
    \includegraphics[width=0.98\columnwidth]{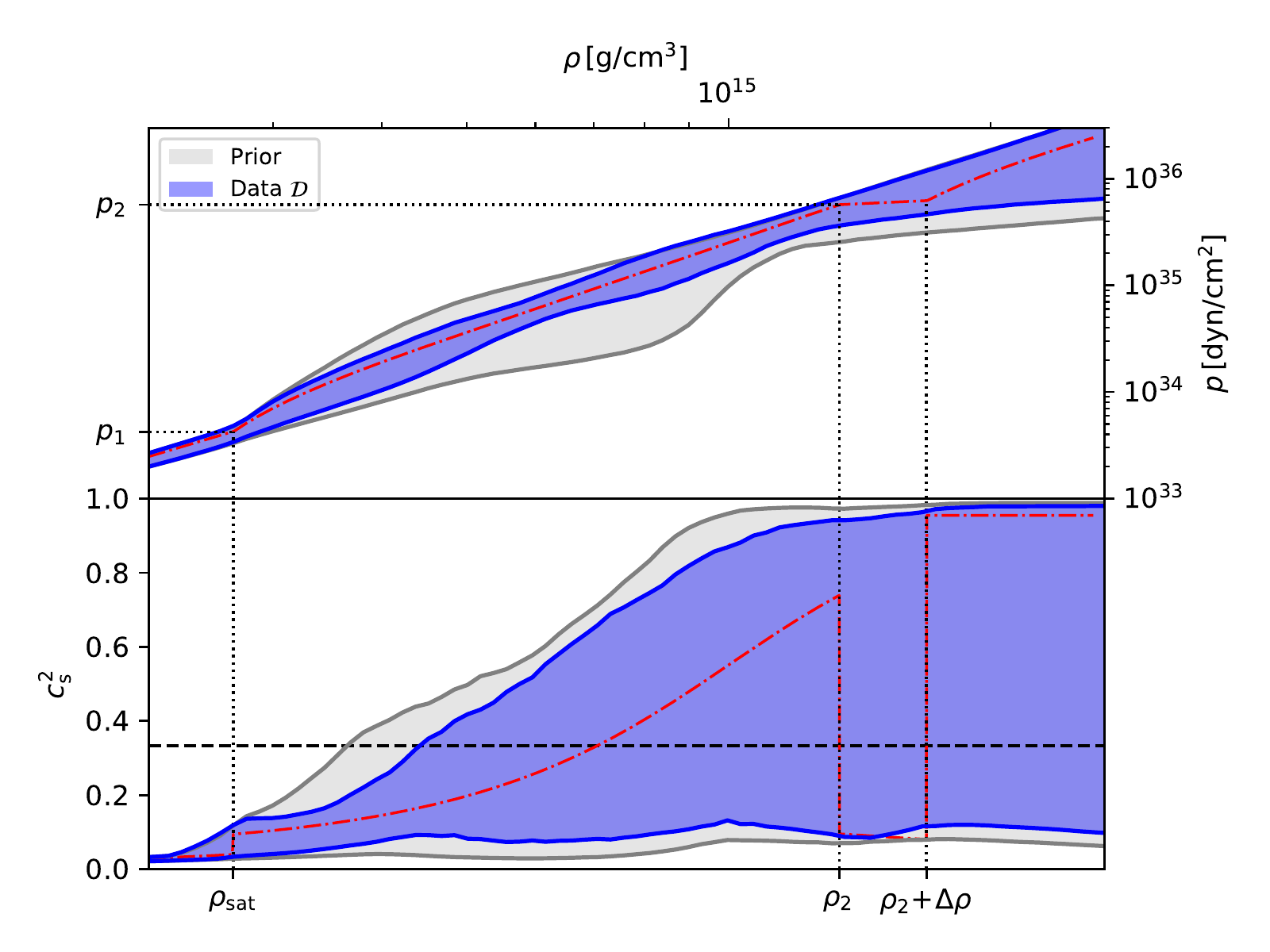}
    \includegraphics[width=0.98\columnwidth]{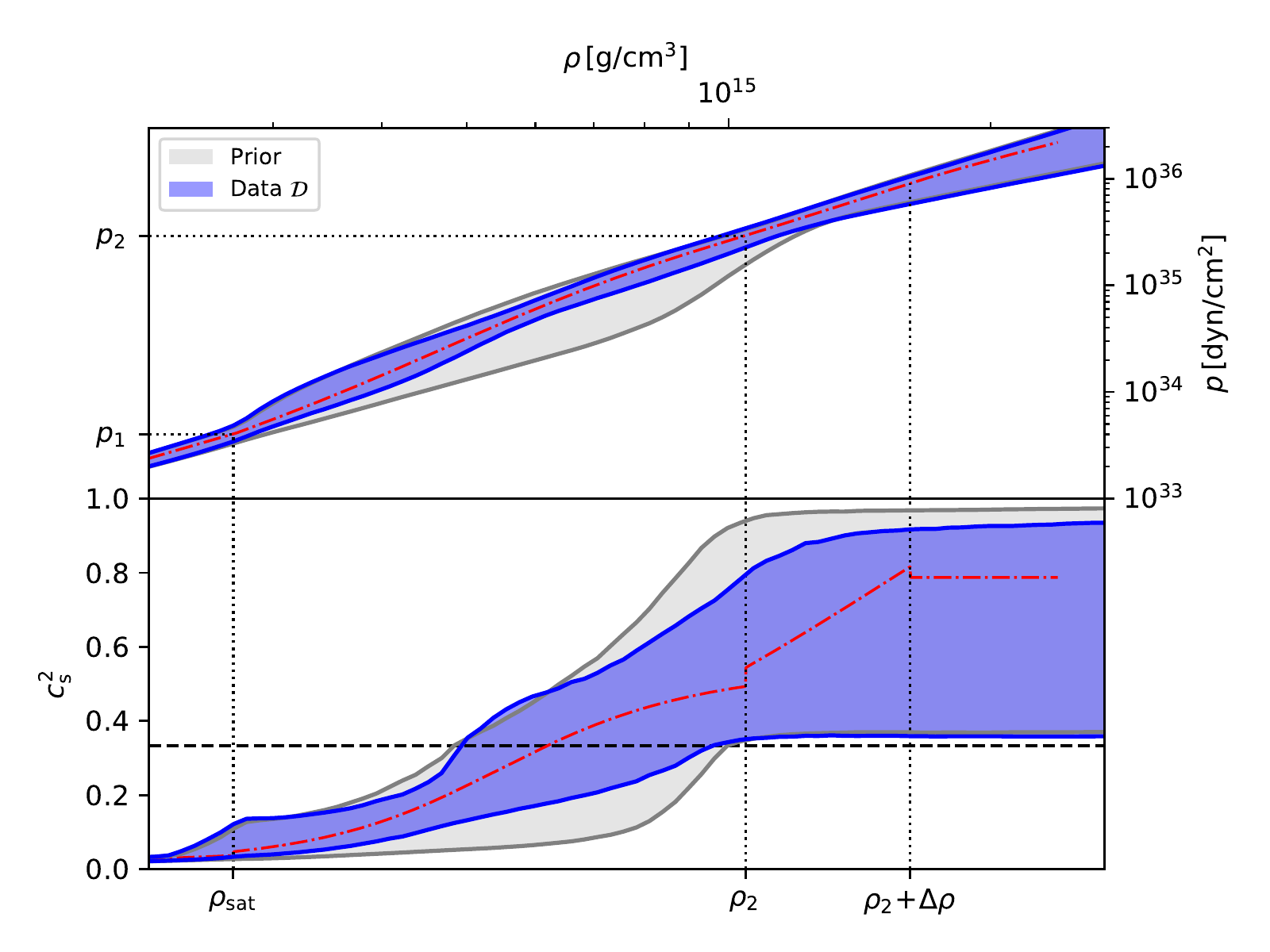}
    \caption{The 90\% uncertainties of the rest-mass density versus the pressure ($\rho-p$) and the rest-mass density versus the squared speed of sound ($\rho-c_{\rm s}^2$) relations. The left (right) panel is the case of the PT (NPT) model. The gray and blue regions represent, respectively, the prior and the posterior constrained with the GW data and the $M-R$ measurements. The red curves are representative EoSs constructed using the associated models, while the horizontal dashed black lines in the bottom panels represent the asymptotic limit $c_{\rm s}^2=1/3$.}
    \label{fig:prhocs}
    \hfill
\end{figure*}

We choose the ranges for the parameters $\vec{\theta}_{\rm EOS}=\{p_1, v_0, v_1, v_2, \rho_2, \Delta \rho, \Gamma_{\rm m}, c_{\rm q}^2\}$ with $p_1 \in[3.12, 4.70]\times10^{33}\,{\rm dyn\,cm^{-2}}$, $v_0 \in[0.4, 4.1]$, $v_1 \in[-3.9, 2.7]$, $v_2 \in[-1.2, 0.8]$, $\rho_2 \in[1, 5]\,\rho_{\rm sat}$, $\Delta \rho \in[0.01, 3.0]\,\rho_{\rm sat}$, $\Gamma_{\rm m} \in[0.01, 1.4]$ (PT model; $\Gamma_{\rm m} \in[1.4, 10]$ for NPT model)\footnote{\citet{2019PhRvD..99j3009M} adopted $\Gamma_{\rm m} = 1.03$ for their Gibbs model. As stressed in the literature \citep{2018PhRvC..97d5802A}, the mixed phase is still uncertain, and therefore we take a wider range for $\Gamma_{\rm m}$. The certain boundary of $\Gamma_{\rm m}$ for phase transition and no phase transition is hard to be determined and is beyond our current research. The limit of 1.4 is only an empirical choice motivated by the fitting results shown in the left panel of Fig.~\ref{fig:fiteos}, where we find that almost all the EoSs have adiabatic indices larger than 1.4 if there is no phase transition.}, and $c_{\rm q}^2 \in[1/3, 1]$, where the ranges of parameters $v_k$ are determined by fitting the theoretical EoSs with the causal spectral representation method. The EoS tables are adopted from the supplementary material of \citet{2016PhRvC..94c5804F} and the appendix of \citet{2020PhRvD.101f3029S}, which include a wide variety of EoS models (some associated references are Refs.~\citep{2003PhRvC..68c1304A, 2005PhRvC..72a4310A, 2010PhRvC..81c4323A, 2014ApJS..214...22B, 1989PhRvC..40.2834B, 1997NuPhA.627..710C, 2016PhRvC..94c5804F, 1986PhRvC..33..335F, 2004NuPhA.732...24G, 1991PhRvL..67.2414G, 2010PhRvC..82c5804G, 2013PhRvC..88b4308G, 2014MNRAS.439..318G, 2006PhRvD..73b4021L, 1997PhRvC..55..540L, 2005PhRvC..71b4312L, 1987PhLB..199..469M, 1995NuPhA.584..467R, 1999PhRvC..60a4316R, 1994NuPhA.579..557S}). We fit both $\rho-\Gamma$ and $\rho-p$ relations up to $5\rho_{\rm sat}$ (as shown in Fig.~\ref{fig:fiteos}), and find that most of the EoSs can be well fitted (within 10\% relative uncertainties) for the $\rho-p$ relation. But for the $\rho-\Gamma$ relation, some EoSs (e.g., the ones with $\sigma$ meson) are hard to be resembled, and the uncertainties can sometimes reach $\sim20\%$. These may be resolved by increasing the number of expanded terms in Eq.~(\ref{eq:Upsilon}), but do not significantly influence our results based on the quality of current observation data. We then select very wide parameter boundaries from the fitting coefficients, and the parameter space can encompass a group of widely-adopted candidate EoSs. Additionally, all of the parametrized EoSs satisfy the following conditions: 1) causality constraint, 2) thermal stability $de/dp>0$, 3) $\Gamma \in [1.4, 10]$ when extending the causal spectral representation to $5\rho_{\rm sat}$, 4) the pressure at $1.85\rho_{\rm sat}$ should exceed $1.21\times10^{34}\,{\rm dyn\,cm^{-2}}$ \citep{2016ApJ...820...28O}, 5) maximum central density of nonrotating NS should exceed $\rho_2$ for PT model, and 6) maximum mass limits $M_{\rm TOV}\in[2.04, 2.3]\,M_\odot$. The left boundary is the 68.3\% lower limit of {PSR J0740+6620}'s mass measurement \citep{2020NatAs...4...72C}, and the upper bound is chosen based on the constraints from the multimessenger analyses of GW170817/GRB 170817A/AT2017gfo \citep{2018ApJ...852L..25R, 2018PhRvD..97b1501R, 2019PhRvD.100b3015S, 2020PhRvD.101f3029S}. The representative EoSs and priors (considering all of the constraints) constructed from our models are shown in Figs.~\ref{fig:prhocs}-\ref{fig:poshigh}. We also investigate the influence of different constraints on our priors. We notice that the EoS parameters for $\rho<\rho_2$ are shaped mostly by the reasonable constraint of $\Gamma$. The causality constraint plays an important role in excluding the very high $\Gamma_{\rm m}$ and $\Delta \rho$ parameter space for the NPT model, and the constraint (5) favors lower values of $\rho_2$ and $\Delta \rho$ for the PT model (without this constraint, both parameters are more uniformly distributed). When we include the $M_{\rm TOV}$ constraint, $\rho_2$ and $\Delta \rho$ are further lowered, but other EoS parameters are not sensitive to it.

\section{Observation Data}\label{sec:data}

One of the informative constraints on the EoS is from the mass-radius ($M-R$) of NSs determined by the traditional spectroscopic measurements or the pulse profile modeling method (see \citet{2016ARA&A..54..401O} for a recent review). The x-ray observations of the low-mass x-ray binaries during quiescence, or those with thermonuclear bursts, have provided us $M-R$ measurements, which, however likely, still suffer from relatively large statistical or systematic uncertainties \citep{2019ApJ...887L..24M, 2016ARA&A..54..401O} and we hence do not take into account almost any of this data. Recently, thanks to the excellent performance of {\it NICER}, the mass and radius of {PSR J0030+0451} were measured with unprecedented precision \citep{ 2019ApJ...887L..24M, 2019ApJ...887L..21R} using the pulse profile modeling method. The radius measurement of {PSR J0437-4715} (this object has a mass of around $1.44\,M_\odot$ that was determined by the reliable timing analyses \citep{2016MNRAS.455.1751R}, and is one of the best targets for {\it NICER}) has been updated in \citet{2019MNRAS.490.5848G}, which will be directly tested by the dedicated {\it NICER} observations in the near future. Via the direct atmosphere-model fits to the time-evolving x-ray burst spectra \citep{2017A&A...608A..31N}, the $M-R$ measurements of {4U 1702-429} were obtained with (significantly) smaller uncertainties in comparison with the sources measured in other indirect ways \citep{2020NatPh..16..907A}. Meanwhile, the induced tidal effects of two tightly interacting NSs can be encoded in the GW stain data. And the remarkable observations of the binary NS merger event GW170817 by LIGO/Virgo detectors \citep{2017PhRvL.119p1101A} have also provided us a novel probe of the EoS \citep{2018PhRvL.121p1101A, 2018PhRvL.120q2703A, 2018PhRvL.121i1102D}. To take advantage of joint analysis of the multimessenger data of NSs, we adopt the data set $\mathcal{D}$, which includes strain data of GW170817 and $M-R$ measurements of {PSR J0030+0451}, {PSR J0437-4715}, and {4U 1702-429} to perform the Bayesian inference.

\section{Bayesian Inference}\label{sec:bayesian}

Assuming that compact stars share the same EoS, we take the likelihood
\begin{equation}\label{eq:likelihood}
    \mathcal{L}=\mathcal{L}_{\rm GW}(d\mid\vec{\theta}_{\rm GW})\times \prod_{i} \mathcal{P}_i(M(\vec{\theta}_{\rm EOS}, h_i), R(\vec{\theta}_{\rm EOS}, h_i))
\end{equation}
to constrain the parameters $\vec{\theta}_{\rm EOS}$ that characterize the ultra dense matter EoS \citep{2020ApJ...888...45T} by performing Bayesian inference with {\sc Bilby} \citep{2019ascl.soft01011A} and {\sc dynesty} \citep{2020MNRAS.493.3132S} as well as {\sc PyMultiNest} \citep{2016ascl.soft06005B} packages. For the $M-R$ observations of {PSR J0030+0451} by {\it NICER} \citep{2019ApJ...887L..24M, 2019ApJ...887L..21R} and {4U 1702-429} \citep{2017A&A...608A..31N}, we use the posterior samples ($\vec{S}$) to construct the kernel density estimate (KDE) as $\mathcal{P}_i(M,R)={\rm KDE}(M,R\mid \vec{S})$ \citep{2020ApJ...892...55J}. However, for {PSR J0437-4715} we approximate the $M-R$ measurements by the products of two KDEs, i.e., $\mathcal{P}_i(M,R)={\rm KDE}(M\mid \vec{S}_{\rm M})\times{\rm KDE}(R\mid \vec{S}_{\rm R})$, where $\vec{S}_{\rm M}$ and $\vec{S}_{\rm R}$ are posterior samples of mass and radius \citep{2016MNRAS.455.1751R, 2019MNRAS.490.5848G}. Each pair of $(M,R)$ is calculated by varying the central enthalpy $h_i$ in the range of [0.06, 0.6]. Besides, the contribution of GW data to the likelihood is determined by its strain data and power spectral densities (detailed processing follows \citep{2020ApJ...888...45T}), waveform models (e.g., {\sc IMRPhenomD\_NRTidalv2}, \citep{2019PhRvD.100d4003D}) as well as the corresponding parameters $\vec{\theta}_{\rm GW}$. We fix the source location of GW170817 to the known position (R.A.=$197.450374^{\circ}$, decl.=$-23.381495^{\circ}$, $z$=0.0099) as determined by electromagnetic observations \citep{2017ApJ...848L..12A, 2017ApJ...848L..28L}. To break the degeneracy between component masses and improve the efficiency in nest sampling, the chirp mass $\mathcal{M}_{\rm c}$ and mass ratio $q$ are sampled instead of $m_{1,2}$. Thus the GW parameters of the marginalized-phase likelihood are $\vec{\theta}_{\rm GW} = \{\Lambda_1(m_1^{\rm src},\vec{\theta}_{\rm EOS}),\Lambda_2(m_2^{\rm src},\vec{\theta}_{\rm EOS})\} \cup \{\mathcal{M}_{\rm c}, q, \chi_{\rm 1z}, \chi_{\rm 2z}, \theta_{\rm JN}, t_{\rm c}, \Psi\}$, where $\Lambda_{1,2}$ are dimensionless tidal deformabilities that are mapped from source frame masses using EoS parameters. Meanwhile, the priors of $\mathcal{M}_{\rm c}$, $q$ are given by $\mathcal{P}(\mathcal{M}_{\rm c},q)\propto \mathcal{M}_{\rm c}(1+q)^{2/5}q^{-6/5}$ ($\mathcal{M}_{\rm c}\in[0.87, 1.74]\,M_\odot$, $q\in[0.5, 1]$) and the additional constraints $m_{1,2}\in[1, 2]\,M_\odot$, which yields uniform distribution in the $m_1-m_2$ plane. Finally, an aligned low-spin prior is assigned to $\chi_{\rm 1z}$ and $\chi_{\rm 2z}$, while $\sin{\!(\theta_{\rm JN})}$ and other parameters (e.g., $\vec{\theta}_{\rm EOS}$, $h_i$, $t_{\rm c}$, and $\Psi$) are uniformly distributed in their domains.

\section{Results}\label{sec:results}

\begin{figure*}
    \centering
    \subfigure[]{\label{fig:postlowparam}
        \begin{minipage}[t]{0.98\columnwidth}
            \includegraphics[width=1.0\columnwidth]{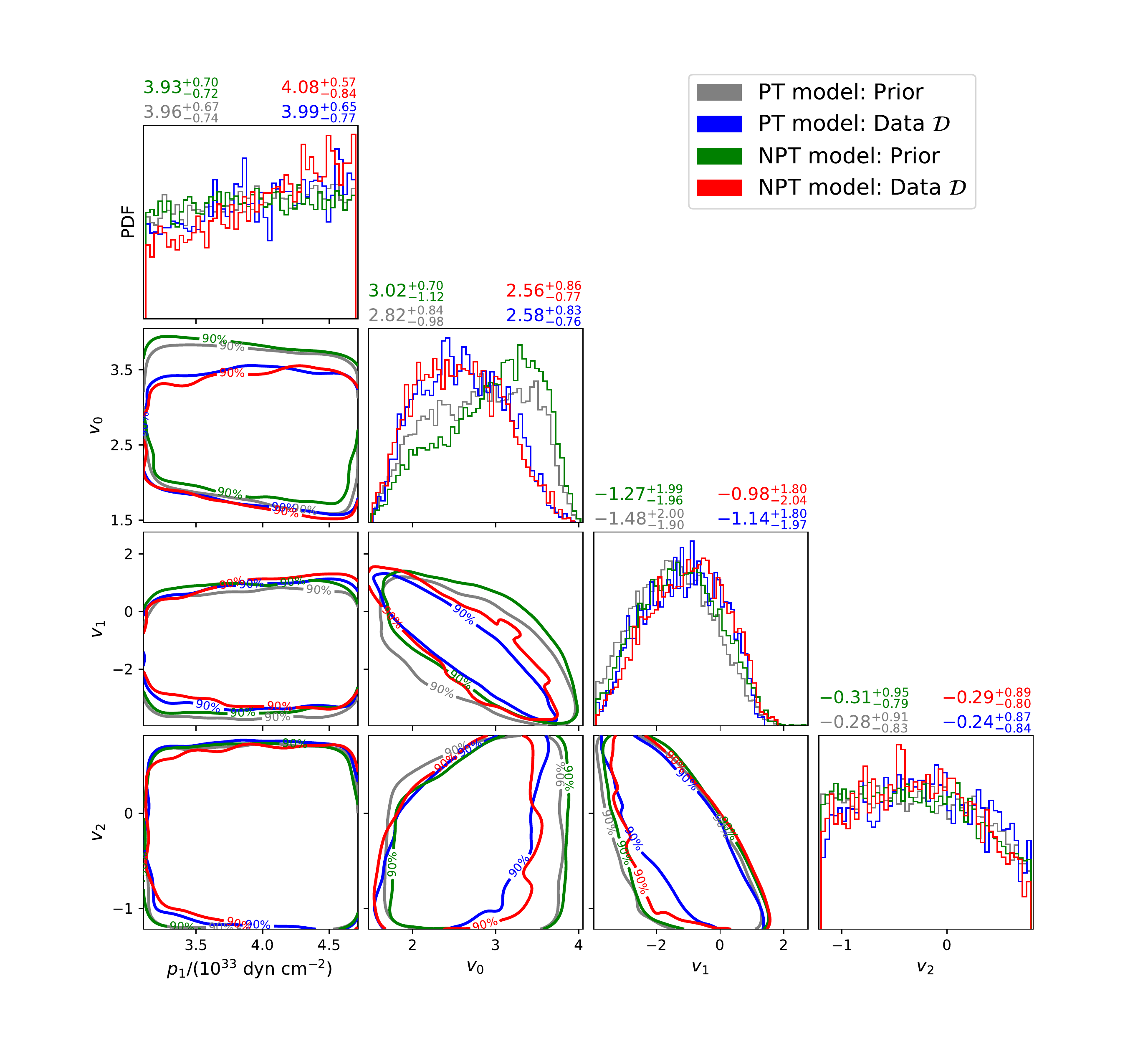}
        \end{minipage}}
    \subfigure[]{\label{fig:postbulkprop}
        \begin{minipage}[t]{0.98\columnwidth}
            \includegraphics[width=1.0\columnwidth]{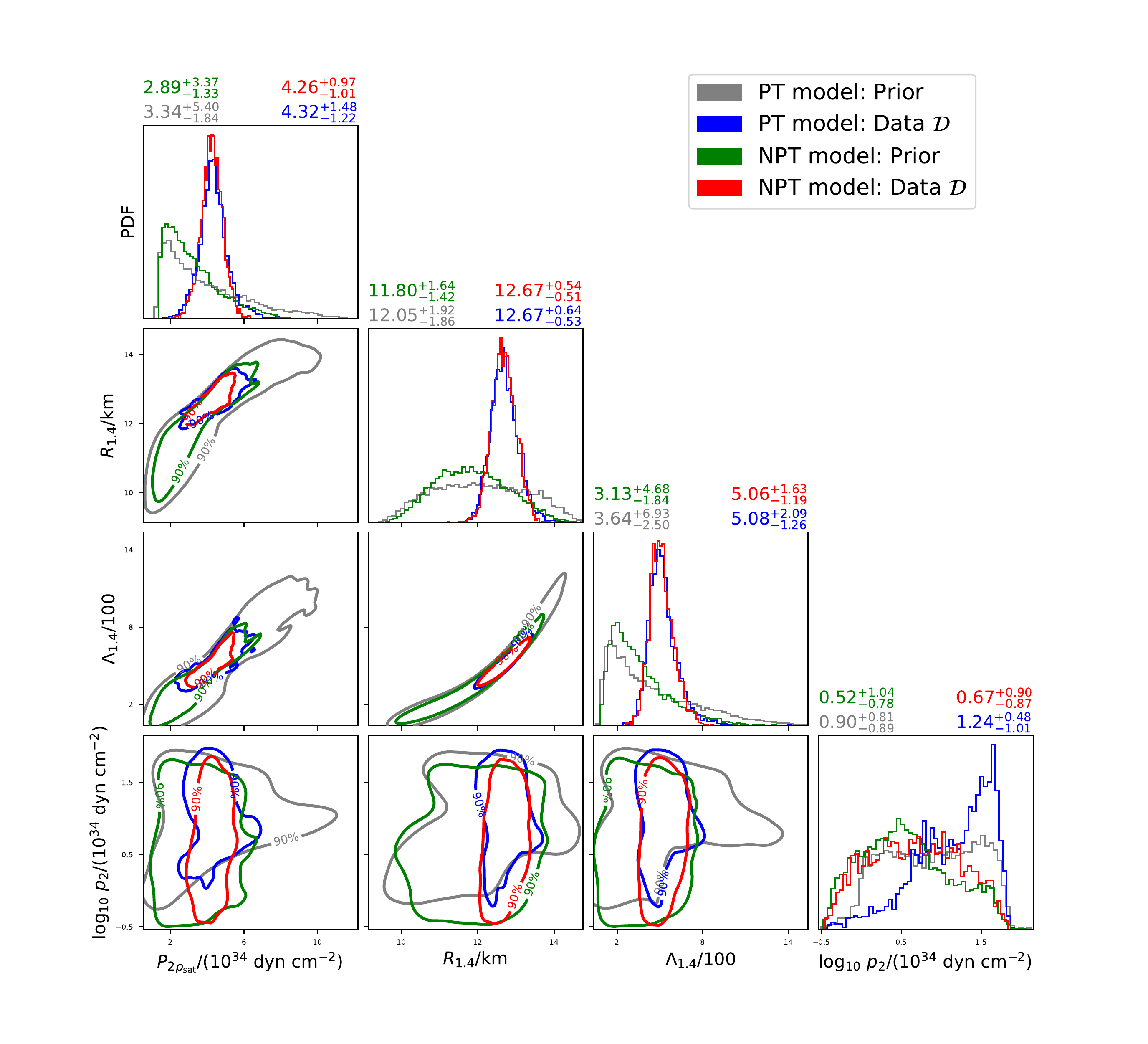}
        \end{minipage}}
    \caption{Distributions of priors (represented by the gray and green colors) and posteriors (represented by the blue and red colors) of the parameters $\{p_1, v_0, v_1, v_2\}$ (left panel), and some inferred microscopic quantities and bulk properties of NS $\{P_{2\rho_{\rm sat}}, R_{1.4}, \Lambda_{1.4}, p_2\}$ (right panel) for PT and NPT models. The values above the diagonal corner plots represent the 90\% credible intervals.}
    \label{fig:postlow}
    \hfill
\end{figure*}

\begin{figure*}
    \centering
    \subfigure[]{\label{fig:poshighpt}
        \begin{minipage}[t]{0.98\columnwidth}
            \includegraphics[width=1.0\columnwidth]{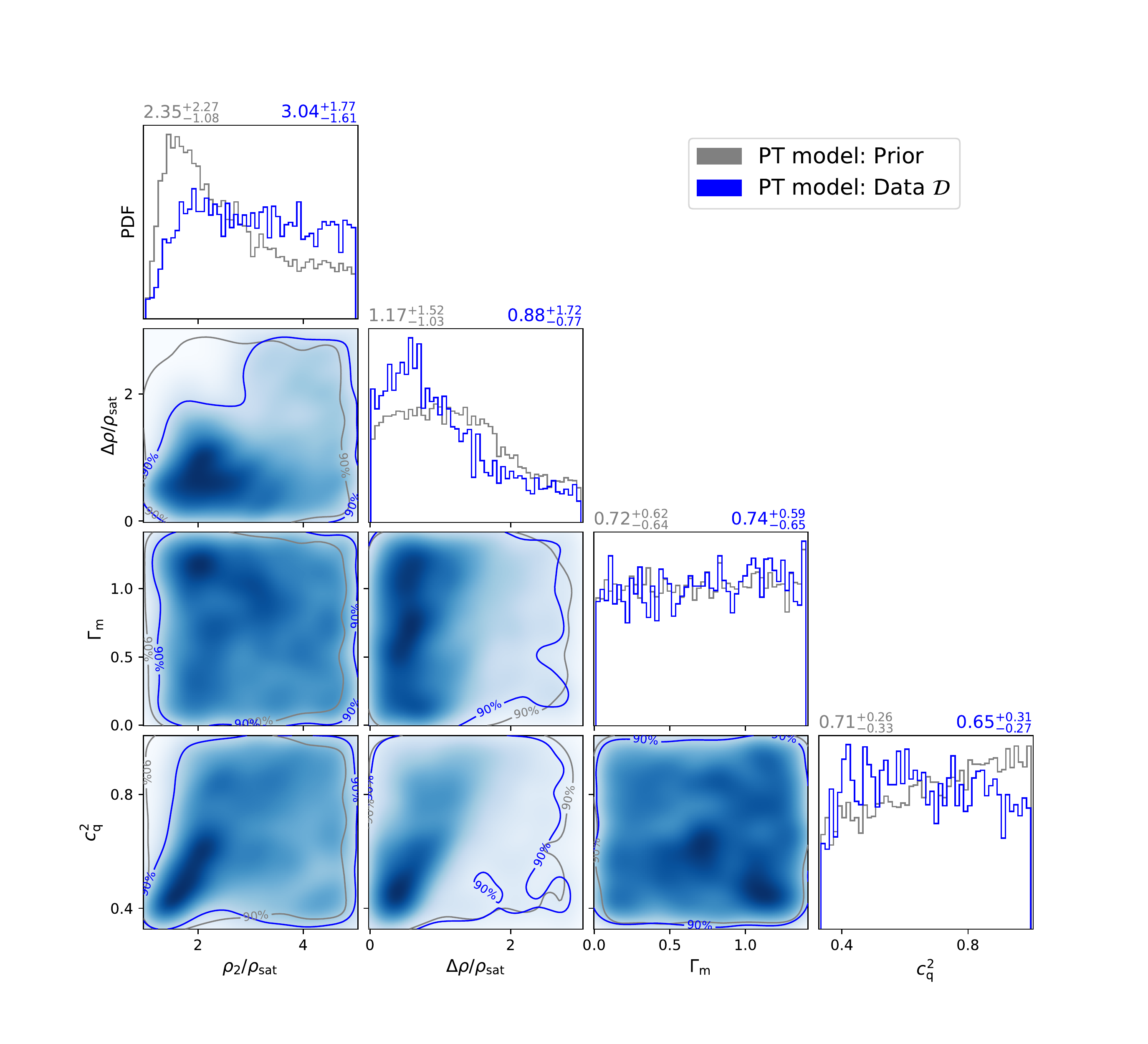}
        \end{minipage}}
    \subfigure[]{\label{fig:poshighnpt}
        \begin{minipage}[t]{0.98\columnwidth}
            \includegraphics[width=1.0\columnwidth]{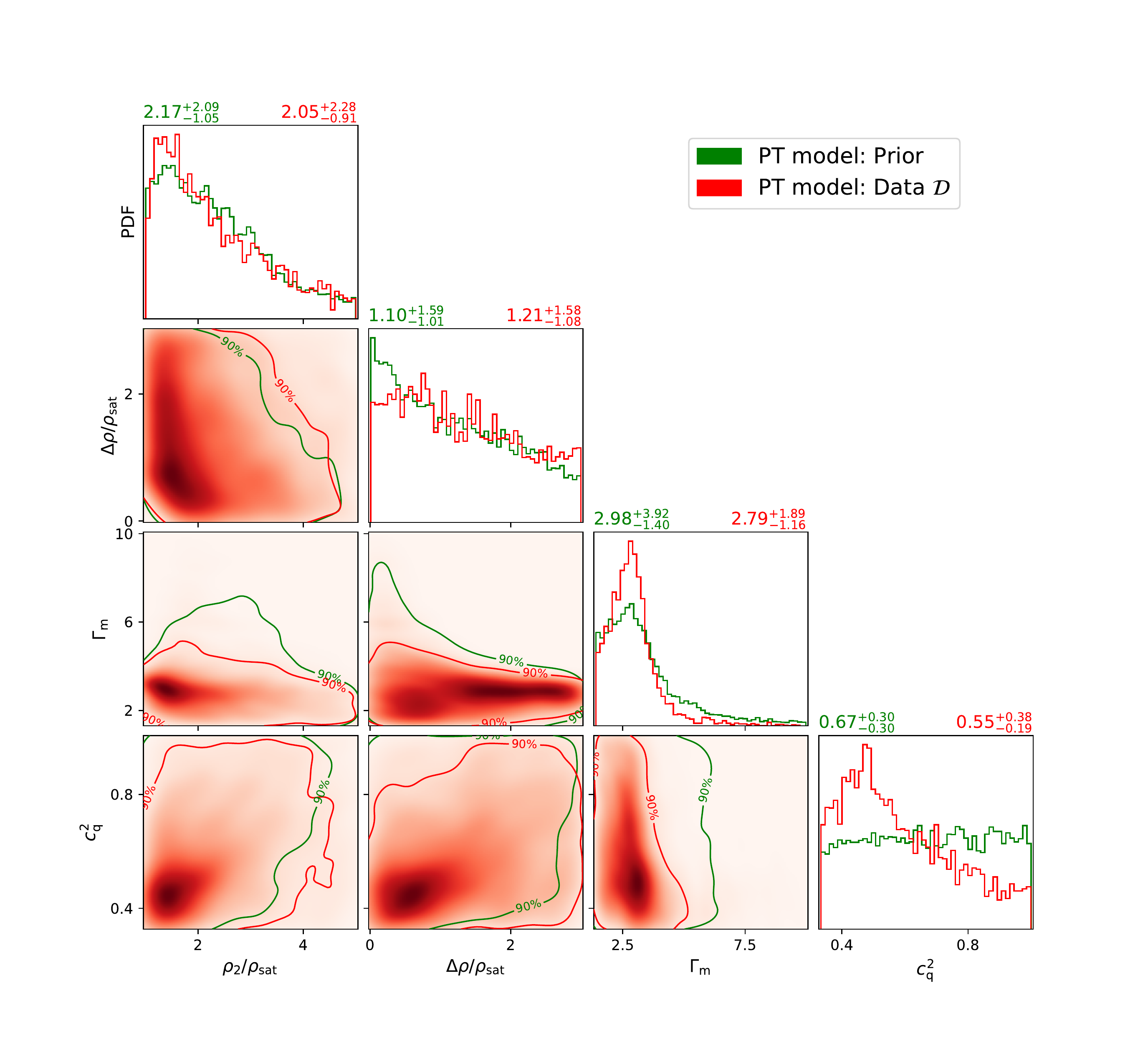}
        \end{minipage}}
    \caption{Distributions of priors (represented by the gray and green colors) and posteriors (represented by the blue and red colors) of the parameters $\{\rho_2, \Delta \rho, \Gamma_{\rm m}, c_{\rm q}^2\}$ for PT (left panel) and NPT (right panel) models. The values above the diagonal corner plots represent the 90\% credible intervals.}
    \label{fig:poshigh}
    \hfill
\end{figure*}

\begin{figure*}
    \centering
    \includegraphics[width=0.98\columnwidth]{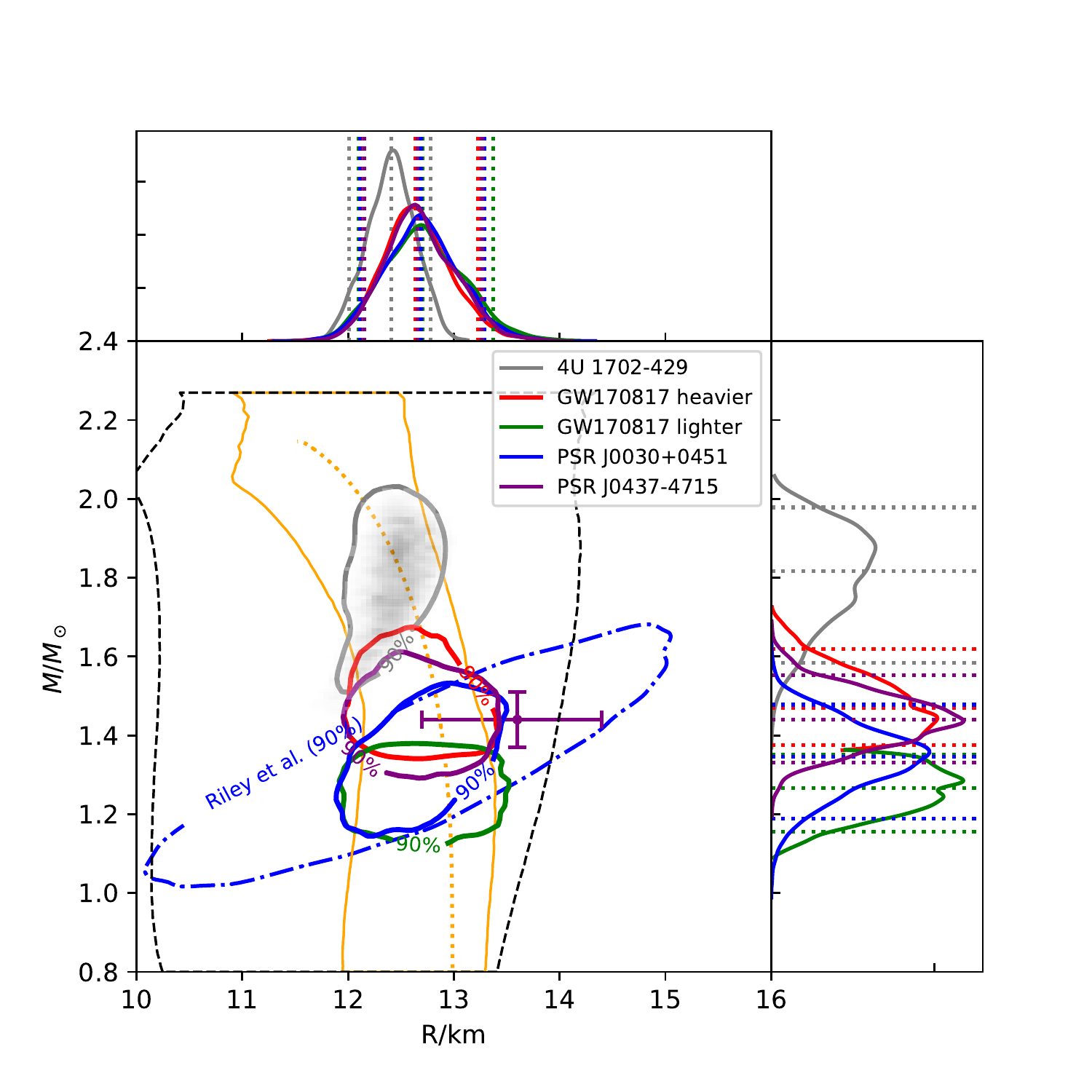}
    \includegraphics[width=0.98\columnwidth]{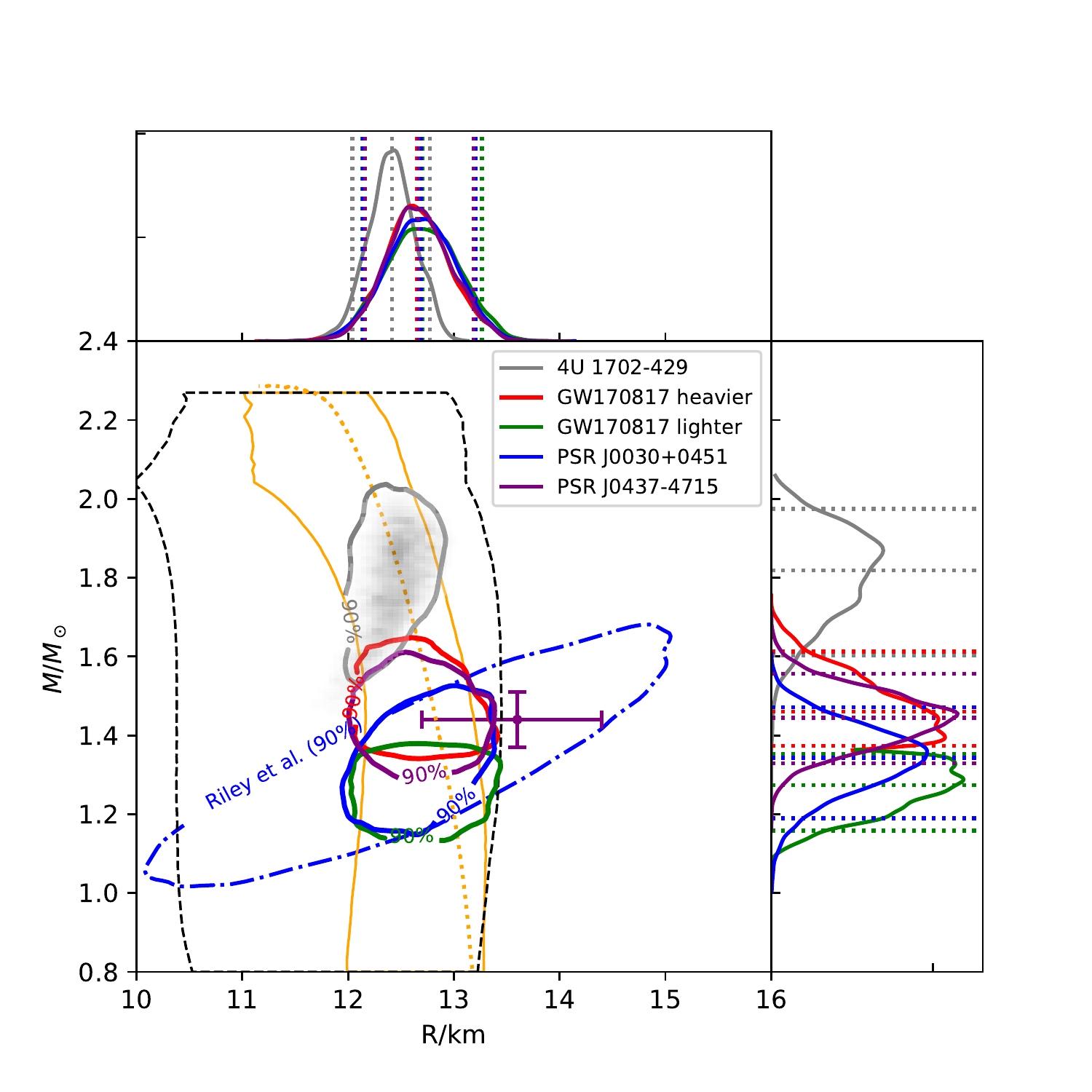}
    \caption{Posterior $M-R$ distributions for PT (left panel) and NPT (right panel) models. The black dashed line and orange solid line denote respectively the 90\% uncertainty region of $M-R$ relations of the prior and the posterior obtained with data set $\mathcal{D}$. The $M-R$ measurements of {PSR J0030+0451}, {PSR J0437-4715}, and {4U 1702-429} are represented by the blue dot-dashed contour, purple error bar, and gray density region, respectively. The associated reconstructed $M-R$ of these sources are represented by the colored solid contours.}
    \label{fig:mr}
    \hfill
\end{figure*}

Our results show that adopting $M-R$ measurements of {PSR J0030+0451} from \citet{2019ApJ...887L..21R} or \citet{2019ApJ...887L..24M} yield rather similar posterior distributions, so we only present the results based on the former. The EoS parameters ($\{p_1, v_0, v_1, v_2\}$) that govern the relatively low density region are loosely constrained by the observation data, except for $v_0$ which slightly deviates from the prior, as shown in Fig.~\ref{fig:postlowparam}. The inferred NS's bulk properties $R_{1.4}$ and $\Lambda_{1.4}$ are constrained to $R_{1.4}=12.67_{-0.53}^{+0.64}\,{\rm km}$ and $\Lambda_{1.4}=508_{-126}^{+209}$ ($R_{1.4}=12.67_{-0.51}^{+0.54}\,{\rm km}$ and $\Lambda_{1.4}=506_{-119}^{+163}$) for the PT (NPT) model, which are strongly correlated with each other \citep{2018PhRvC..98c5804M} since these properties are mainly determined by the pressure at around $2\rho_{\rm sat}$ \citep{2016PhR...621..127L} that is well constrained by the data [see Fig.~\ref{fig:postbulkprop}]. Meanwhile, the $\rho-p$ relations of EoS in the relatively low density region of Fig.~\ref{fig:prhocs} are also tightened. This is understandable, because most sources we adopted in the analyses have masses centered in the low mass region (as shown in Fig.~\ref{fig:mr}) and thus have relatively low central densities. However, current data are still hard to give insight into the sound velocity property of dense matter, and the $\rho-c_{\rm s}^2$ relation remains less constrained compared with the $\rho-p$ relation.

We compare the results of the PT and NPT models regarding the EoS parameters above the dividing density $\rho_2$. In the case of the PT model, we find that the priors have already put strong constraints on the parameter space, and the joint analysis of GW170817 and the $M-R$ measurements of the three sources extracts little information about the parameters of $\Delta \rho$, $\Gamma_{\rm m}$, and $c_{\rm q}^2$. However, there are still some noticeable differences compared with the priors. The $\rho_2-\Delta \rho$ corner plots in Fig.~\ref{fig:poshighpt} indicate that strong phase transition at the low density region is not favored, and the $1\sigma$ lower bound of $\rho_2$ is constrained to $1.84\rho_{\rm sat}$. With a different approach, \citet{2020ApJ...894L...8C} concluded that strong phase transition below $1.7\rho_{\rm sat}$ ($1\sigma$ level) was ruled out. The disfavor of low transition density may be explained by the fact that if phase transition occurs at very low density with a long platform of pressure, it is relatively difficult for the pressure at around $2\rho_{\rm sat}$ to achieve the higher values favored by the data [as shown in Fig.~\ref{fig:postbulkprop}]. The posterior distribution of $p_2$ (the corresponding pressure at $\rho_2$) of the PT model in Fig.~\ref{fig:postbulkprop} also shows the disfavor of very low values compared with its prior. Thus we put a $90\%$ lower bound for the transition pressure $p_2>1.70\times 10^{34}\,{\rm dyn\,cm^{-2}}$. In the case of the NPT model, the posterior distributions are similar to the priors except the parameter $\Gamma_{\rm m}$ and $c_{\rm q}^2$ which dominate the behavior of EoS at high densities. For these two parameters, we find that larger values are strongly disfavored by the data, which lead to a $90\%$ upper limit of $\Gamma_{\rm m}<4.68$.

We also investigate the impact of adopting the subsets of our observation data. The bulk properties of NSs differed when we only use the GW data and the $M-R$ measurements of {PSR J0030+0451}, which gives $R_{1.4}=11.5_{-1.0}^{+1.5}\,{\rm km}$ and $\Lambda_{1.4}=250_{-115}^{+350}$ (the results from the PT and NPT are similar, so we only report averaged values). The analysis of all the $M-R$ measurements gives $R_{1.4}=12.9_{-0.7}^{+1.0}\,{\rm km}$ and $\Lambda_{1.4}=570_{-170}^{+490}$. As for the EoS parameters above the dividing density $\rho_2$, using different groupings of the measurements yields posterior distributions that are more similar to the priors compared to analyzing the whole data set, but they also show disfavor of strong phase transition occurring at relatively low densities. The reconstructed $M-R$ distributions indicate that the second peak of the combined tidal parameter ($\tilde{\Lambda}$) of GW170817 is favored, and the radius of {PSR J0437-4715} is constrained to $R\!\approx\!12.7\!\pm\!0.6\,{\rm km}$, which is in the low range of the evaluation by \citet{2019MNRAS.490.5848G} (as shown in Fig.~\ref{fig:mr}). The $M-R$ measurements of {PSR J0030+0451} and {4U 1702-429} are well reproduced. For both models, the joint analysis of data set $\mathcal{D}$ largely narrows down the $90\%$ uncertainties of $M-R$ curves compared with the priors. However, it seems to be not obvious for exhibiting distinct $M-R$ characteristics between PT and NPT models, because the observable EoS feature may be smeared out by the mixed phase. Meanwhile, the evidences of PT and NPT models are comparable, with a Bayes factor of $\mathcal{B}_{\rm NPT}^{\rm PT}\sim1.2$. Therefore, current data are not informative enough to neither support nor rule out phase transition. Future radius measurements of massive NS may be promising to probe such transition by the joint analysis with $M_{\rm TOV}$ and $R_{\sim1.4}$ constraints \citep{2020ApJ...899..164H}.

\section{Discussion and Summary}\label{sec:summary}

We have constructed two empirical models, i.e., PT and NPT models, with different constraints and priors. We then performed the Bayesian parameter inference with the GW data (GW170817) and $M-R$ observations ({PSR J0030+0451}, {PSR J0437-4715}, and {4U 1702-429}) using the phenomenologically constructed EoS models to search for potential first-order phase transition. We find that the bulk properties of NSs, i.e., the radius and tidal deformability of canonical $1.4\,M_\odot$ NS, are well constrained to $R_{1.4}=12.67_{-0.53}^{+0.64}\,{\rm km}$ and $\Lambda_{1.4}=508_{-126}^{+209}$ ($R_{1.4}=12.67_{-0.51}^{+0.54}\,{\rm km}$ and $\Lambda_{1.4}=506_{-119}^{+163}$) for PT (NPT) models. We also find that current observation data are still too uninformative to decide whether phase transition exists in NSs, because the evidence for both models are comparable and the parameters $\vec{\theta}_{\rm EOS}$ are dominated by the priors. However, when assuming if phase transition is really present in NSs, we conclude that strong phase transition at low densities is not favored by the observation data. Note that our PT model is a more general case incorporating both Maxwell-like and Gibbs-like phase transitions, but a masquerade problem will appear for Gibbs-like EoSs, which may make their macroscopic structure properties ($M-R$ or $M-\Lambda$ relations) hard to be distinguished from purely nucleonic EoSs \citep{2005ApJ...629..969A}. Benefiting from the dedicated observations by {\it NICER}, unprecedentedly precise $M-R$ measurements for massive NSs (e.g., {PSR J0740+6620} and {PSR J1614-2230} \citep{2019ApJ...887L..27G}) as well as {PSR J0437-4715} will be available in the future. Hence, with our phenomenological parametrization model that is more generic and flexible and able to resemble various theoretical EoS models, the existence of phase transition or no phase transition will be further probed, and then we will be able to shed valuable light on the dense matter in the core of NSs.
\\

\section{Acknowledgments}
We appreciate the anonymous referees for their very helpful suggestions. We also thank D. Gonz{\'a}lez-Caniulef and J. N{\"a}ttil{\"a} for providing us posterior samples of mass-radius measurements. This work was supported in part by NSFC under Grants No. 11921003, No. U1738126, No. 11933010, and No. 12073080, as well as the Chinese Academy of Sciences via the Strategic Priority Research Program (Grant No. XDB23040000) and the Key Research Program of Frontier Sciences (No. QYZDJ-SSW-SYS024).

\bibliography{ms.bbl}

\end{document}